\documentclass[12pt]{emulateapj}
\def\Meszaros{M\'esz\'aros~}

\begin{document}

\title{On the origin and survival of UHE cosmic-ray
nuclei in GRBs and hypernovae}

\author{Xiang-Yu  Wang\altaffilmark{1,2,4},  Soebur Razzaque\altaffilmark{1,3,4} and  Peter
M\'esz\'aros\altaffilmark{1,3,4}}

\altaffiltext{1}{Department of Astronomy and Astrophysics,
Pennsylvania State University, University Park, PA 16802, USA}
\altaffiltext{2}{Department of Astronomy, Nanjing University,
Nanjing 210093, China} \altaffiltext{3}{Department of Physics,
Pennsylvania State University, University Park, PA 16802, USA}
\altaffiltext{4} {Center for Particle Astrophysics, Pennsylvania
State University, University Park, PA 16802, USA}

\begin{abstract}
The chemical composition of the ultra-high-energy (UHE) cosmic
rays serves as an important clue for their origin. Recent
measurements of the elongation rates by the Pierre Auger
Observatory  hint at the possible presence of heavy or
intermediate mass nuclei in the UHE cosmic rays. Gamma-ray bursts
(GRBs) and hypernovae have been suggested as possible sources of
the UHE cosmic rays. Here we derive the constraints on the
physical conditions under which UHE heavy nuclei,  if they are
accelerated in these sources, can survive in their intense photon
fields.  We find that in the GRB external shock and in the
hypernova scenarios, UHE nuclei can easily survive
photo-disintegration. In the GRB internal shock scenario, UHE
nuclei can also survive, provided the dissipation radius and/or
the bulk Lorentz factor of the relativistic outflow are relatively
large, or if the low-energy self-absorption break in the photon
spectrum of the prompt emission occurs above several KeV. In
internal shocks and in the other scenarios, intermediate-mass UHE
nuclei have a higher probability of survival against
photo-disintegration than UHE heavy nuclei such as Fe.
\end{abstract}


\section{Introduction}
Ultra-high-energy (UHE) cosmic rays above $\sim10^{18}$ eV are
thought to be of extra-galactic origin, since charged particles
with such high energies cannot be confined by the magnetic field
of our Galaxy. Galactic cosmic ray accelerators, such as supernova
remnants, are expected to reach their maximum energy below
$\sim10^{18}$ eV. The transition between galactic and
extra-galactic cosmic rays is therefore believed to happen either
at the 'second knee' around $10^{18}$ eV, where the chemical
composition changes significantly as measured by HiRes, followed
by a 'dip' in the spectrum (Berezinsky et al. 2006), or at the
'ankle' around $10^{19}$ eV where the cosmic ray spectrum becomes
flatter.

At these ultra-high energies, the chemical composition is a
subject of debate. It has been speculated that these cosmic rays
are made of pure protons, up to the highest energies. On the other
hand, there are also reasons for favoring a cosmic ray spectrum
dominated by heavy or intermediate mass nuclei at the highest
energies, since according to the Hillas criterion (Hillas 1984)
astrophysical sources are able to accelerate particles up to a
maximum energy proportional to their nuclear charge. Recently, a
mixed composition scenario has been invoked to fit the UHECR
spectrum above $\sim10^{18.5}{\rm eV}$ (Allard et al. 2005). By
studying the details of the development of the resulting air
showers, one can in principle infer the species of the primary
UHECRs, since at a given energy, showers initiated by heavy nuclei
develop higher up in the atmosphere than proton-induced showers.
Both AGASA and HiRes data favor a dominance of light hadrons,
consistent with being pure protons, in the composition of UHECRs
above $10^{19}$ eV (Hughes et al. 2007), which is consistent with
models assuming UHECRs above $10^{18}$ eV are due to
extra-galactic protons. {On the other hand, a recently report of
an analysis of the muon component of UHE air showers in the
Yakutsk data showed a heavy nuclei component in the UHECR spectrum
(Glushkov et al. 2007). }

One of the aims of Pierre Auger Observatory is to study the
composition of UHECRs. It provides more precise measurements of
the depth of the UHECR-initiated shower maximum $X_{\rm max}$ at
extremely high energies ($>10^{19}$ eV), with uncertainties which
are about a factor 4 smaller than those of the best measurement
from the HiRes experiment (Unger et al. 2007). The elongation rate
data presented by the Pierre Auger Observatory team is
satisfactorily represented by a fit containing a break point in
the slope at $10^{18.35}$ eV (Unger et al. 2007). Due to the
uncertainties in the hadronic interactions at the highest
energies, the interpretation of these elongation rate depends on
the hadronic physics models used in the analysis, and is therefore
rather ambiguous at present.  However, regardless of which
interaction models are used, the data appears to require the
presence of a substantial fraction of heavy or intermediate-mass
nuclei in the range of GZK cutoff energies.  A possible conflict
in the chemical composition between the above Auger elongation
results and recent results by the Pierre Auger collaboration
(2007) on large scale spatial correlations is an issue which
remains to be resolved.

The proposed astrophysical source models of UHECRs include AGN
jets (e.g.  Biermann 1987; Berezinsky et al. 2006), intergalactic
accretion shocks (e.g. Inoue, Aharonian \& Sugiyama, 2005), and
gamma-ray bursts (e.g.  Waxman 1995, 2004a; Vietri 1995; Wick et
al. 2004; Dermer \& Atoyan 2006; Murase et al. 2006). Recently, we
have proposed that extra-galactic hypernovae associated with
sub-energetic GRBs are also a probable source for UHECRs, whose
energetics are sufficient large to account for UHECRs from the
second knee and above (Wang et al. 2007). These UHECR accelerators
also accelerate electrons, which produce optical, x-ray or
gamma-ray photons through synchrotron and inverse Compton
emission. Since a sufficiently high density of these low-energy
photons can potentially lead to photo-disintegration of the cosmic
ray heavy nuclei, a natural question is under what conditions do
these sources allow the survival of UHE heavy nuclei from the
acceleration site. In this paper we explore this problem for GRBs
and hypernovae sources{\footnote{The photo-disintegration of UHE
nuclei in AGN scenario has been discussed in Anchordoqui et al.
(2007b)  and Dermer (2007a).}}, since they have the most compact
source sizes and might therefore pose the greatest hurdle for UHE
heavy nuclei survival.

Long-duration GRBs are generally believed to result from the core
collapse of massive stars, direct evidence for which comes from
the spectroscopic identification of bright supernovae in
association with these GRBs (see Woosley \& Bloom 2006 for a
review). The collapse leads to a black hole (or magnetar) plus
accretion disk system. The annihilation of neutrinos or Poynting
flows arising from the inner hot accretion disk produce a high
entropy fireball outflow, which expands and converts its internal
energy into the kinetic energy of a small amount of baryon
material, while propagating through the star along the rotation
axis of the collapsing core. The outflow gets collimated and
finally two highly relativistic jets break out of the stellar
envelope. In the GRB scenario for UHECRs, both internal shocks
(Waxman 1995) and external shocks (Vietri 1995; Wick et al.  2004;
Dermer \& Atoyan 2006) in the jets are suggested to be able to
accelerate baryons to ultra-high energies. Internal shocks develop
due to internal collisions between the shells in the fireball jet
ejecta, while the external shocks occur when the fireball ejecta
interacts with the surrounding interstellar medium (see \Meszaros
2006 and Zhang 2007 for recent reviews). In the internal shock
scenario, viewed as a source of the cosmic ray nuclei, the jets
themselves must contain heavy nuclei, whereas in external shocks
the heavy nuclei could come from the swept-up circumburst material
that the jet is running into, presumably the stellar wind or
interstellar medium. Since the two shocks have quite different
dissipation radii and emission properties, the disintegration
problem of the UHE comic ray nuclei are different, as discussed in
\S\S \ref{sec:int} and \ref{sec:ext} respectively.

Hypernovae are a peculiar type of supernovae with higher ejection
velocities of the remnant shell and generally larger explosion
energies than typical supernovae (Paczy{\'n}ski 1998). The
prototype of hypernovae is SN1998bw, a type Ic supernova
associated with an under-energetic GRB, GRB980425 (Galama et al.
1998). Mildly relativistic ejecta components are inferred to be
present in all three of the well-identified
hypernova/under-energetic GRB systems so far, namely
SN1998bw/GRB980425, SN2003lw/GRB031203 and SN2006aj/GRB060218.
Based on this mildly relativistic ejecta component and the event
rates of these objects, we have shown that hypernovae can
accelerate particles to $\sim Z\times10^{19}$ eV, and the
energetics and occurrence rate is sufficient to account for the
flux of UHECRs above the second knee ($\sim 6\times10^{17}$) eV,
where $Z$ is the nuclear charge of the accelerated particles (Wang
et al. 2007). The particles are accelerated in the hypernova blast
wave formed by the interaction between the hypernova ejecta with
the surrounding stellar wind medium. In \S \ref{sec:hn} we study
whether UHE heavy nuclei accelerated in such hypernovae can survive
their photon environment.

{The goal of this paper is therefore to answer the question, if
UHE heavy nuclei are accelerated in GRBs or hypernovae, what are
the possible origins  of these nuclei, and under what conditions
would they survive their source environment; or conversely, under
what conditions would one expect these sources to accelerate
mainly protons to UHE energies.}

\section{Origin of heavy or intermediate mass nuclei in the sources}
\label{sec:origin}

It is at present unknown whether heavy or intermediate mass nuclei are
present in GRB jets. At the base of the outflow the jets start out as
a hot fireball, within a region of size $10^6-10^7{\rm cm}$ and
temperature $kT=1-10 {\rm MeV}$, where any heavy nuclei will be
photo-disintegrated due to the abundance of photons with energies
comparable to the nuclear binding energy, $\sim10 {\rm MeV}$. The
fireball is thus initially made up of free nucleons, $e^{\pm}$ pairs,
trapped blackbody radiation and magnetic fields.

As the fireball expands and cools, the free nucleons in the jet
may recombine into $\alpha$-particles (Beloborodov 2003), but they
will not form heavy nuclei. However, as the jet burrows through
the stellar core (from inside outward, Fe, O, C cores in
sequence), heavy nuclei from the stellar surroundings could be
entrained into the jet. According to numerical simulations of jet
propagation (Zhang et al. 2003), Kelvin-Helmholtz instabilities
and/or oblique shocks that develop lead to the mixing of
surrounding material into the jet, while the jet is advancing with
a sub-relativistic velocity. Since the temperature of the thermal
photons trapped in the jet decreases significantly as jet expands,
with $T(r)\sim r^{-1}$, these thermal photons are no longer able
to disintegrate the entrained nuclei when the jet has reached the
radius of the stellar Fe core at $\sim 10^9{\rm cm}$. To avoid
spallation of the nuclei due to nucleon-nucleon collisions, the
relative velocity between the jet and the surrounding core
material should be below a critical value $\beta_{\rm
sp}\simeq0.14$ (in units of the speed of light), at which the
relative kinetic energy equals the nuclear binding energy $\sim
10{\rm MeV}$. The calculations show that the jet head moves with a
velocity about $10^9 {\rm cm s^{-1}}$ inside the Fe core at
$\sim10^9 {\rm cm}$ and then the velocity increases as $\sim
r^{1/2}$ (\Meszaros \& Rees 2001). Thus, the Fe and O nuclei
entrained from the surroundings can survive both
photo-disintegration and nuclear spallation.

After the jet head breaks out of the star, in its subsequent
stages the jet continues to escape through the evacuated funnel
cavity through the star, so it accelerates to a relativistic
velocity at radii still inside the star. Thereafter, the
Kelvin-Helmholtz instability which causes the mixing of nuclei
will be suppressed due to the relativistic relative motion (Zhang
et al. 2003). Moreover, since the relative velocity between the
jet and the surrounding star exceeds the critical velocity
$\beta_{sp}$, spallation effects will need to be taken into
account. One finds that the time for spallation
$t_{sp}=1/(\sigma_{sp}n_N c)$ is much shorter than the dynamic
time $r/(\Gamma c)$ when the relativistic jet is inside the
progenitor star, where $n_N=L/({4\pi r^2\Gamma^2 m_N c^3})$ is the
nucleon density of the jet and $\Gamma$ is the Lorentz factor of
jet. This means that even if nuclei get entrained, they will be
spalled into lighter nuclei. Since the break-out time for initial
jet after its inception is about $t_b=10 \,{\rm s}$, long GRBs
with rest-frame durations $t_\gamma \sim2-10$ s will have spent
most of their life inside the star, while the jet head was moving
sub-relativistically, so most of the jet can have become
Fe-enriched before the internal shocks occur. For longer duration
jets, with $t_\gamma\ga 10{\rm s}$, only the first $\sim 10 \,{\rm
s}$ portion of the burst may be Fe-rich, while the rest of the
outburst may consist of pure protons.

The above discussion is relevant to the models that invoke the
composition of the GRB jet as UHECR nuclei, which are accelerated in
internal shocks that are widely assumed to be responsible for the
prompt $\gamma$-ray emission. For GRB external shocks and hypernova
remnant blast wave shocks, the nuclei could be due to the material
swept-up by the shock front. The progenitors of long-duration GRBs and
hypernovae are thought to be Wolf-Rayet stars, as the spectral type of
the discovered supernovae in these events is typically Ic. These stars
are stripped of their hydrogen envelope and sometimes even the helium
envelope. The heavy or intermediate mass UHE nuclei may originate from
the stellar wind of the Wolf-Ryet star. In WC type Wolf-Rayet stars,
the C abundance is $X_{\rm C}=20\%-55\%$ (by mass) and the O abundance
is $X_{\rm O}=5-10\%$ (Crowther et al. 2007). In WO type Wolf-Rayet
stars, O and C abundances are even higher, with $X_{\rm O}=15-25\%$
and $X_{\rm C}=40\%-55\%$ (e.g. Kingsburgh et al.  1995). The
abundance of heavy elements in stellar wind of these Wolf-Rayet stars
is clearly much higher than the solar value.

\section{Disintegration of UHE nuclei}
\label{sec:disint}

The most relevant processes that may prohibit acceleration of heavy
nuclei to UHE at the astrophysical sources are the
photodisintegration, photopion production and nuclear spallation.  The
former two processes are due to interactions of UHE nuclei with
surrounding photons and the latter is due to interactions with other
nuclei or nucleon.  Both the photodisintegration and photopion
processes are dominated by resonance production of either an excited
state of the nuclei or a Delta and subsequent de-excitation or
decay. In case of photodisintegration this dominant channel is called
a giant dipole resonance (GDR) in the energy range $\sim 10-30~{\rm
MeV}$ with a threshold energy of $\sim 10~{\rm MeV}$ per nucleon.  In
case of photopion the threshold energy for pion production is $\simeq
150~{\rm MeV}$.

For an UHE nucleus with Lorentz factor $\gamma_A$ propagating
through an isotropic photon background with differential number
density $n(\varepsilon)$ at energy $\varepsilon$, the
photodisintegration or photopion rate is given by (Stecker 1968)
\begin{equation} t_{\rm
dis,pi}^{-1}=\frac{c}{2\gamma_A^2}\int_{\varepsilon_{th}}^{\infty}
\varepsilon'\sigma_{A,\pi}(\varepsilon')d\varepsilon'
\int_{\varepsilon'/2\gamma_A}^{\infty}\frac{n(\varepsilon)}
{\varepsilon^2} d{\varepsilon} ,
\end{equation}
where $\varepsilon'$ is the photon energy in the rest frame of the
nucleus.  The threshold energy $\varepsilon_{th} = 10$~MeV for
photodisintegration and $m_\pi c^2 + m_\pi^2 c^2/2m_p\simeq145{\rm
MeV}$ for photopion.  The cross-sections for photodisintegration
in the energy range $\varepsilon_{th} < \varepsilon' \lesssim
30$~MeV  with loss of one nucleon can be approximately described
by a Lorentzian form (Puget, Stecker \& Bredekamp 1976;
Anchordoqui et al. 2007a) as
\begin{equation}
\sigma_A(\varepsilon')=\frac{\sigma_{0,A}
{\varepsilon'}^2\Delta_{GDR}^2}{({\varepsilon}_0^2-{\varepsilon'}^2)^2
+{\varepsilon'}^2\Delta_{GDR}^2},
\end{equation}
where $\Delta_{GDR}$ and $\sigma_{0,A}$ are the width and maximum
value of the cross section, $\varepsilon_0$ is the energy at which
the cross section peaks.  Fitted numerical values are
$\sigma_{0,A}=1.45A\times10^{-27}~{\rm cm^2}$,
$\Delta_{GDR}=8~{\rm MeV}$, and $\varepsilon_0=42.65
A^{-0.21}$~MeV for $A>4$ (Karakula \& Tkaczyk 1993).  { Above
$\varepsilon' \sim 30$~MeV and below the pion production threshold
energy, photodisintegration may result in multiple nucleon
emission, although with much lower cross-section. Puget, Stecker
\& Bredekamp (1976) suggested a parametrization such that the
cross-section $\sigma_A(\varepsilon')$ integrated in the range
$\varepsilon_{th} < \varepsilon' \lesssim 30$~MeV and in the range
$30~{\rm MeV} < \varepsilon' \lesssim 150$~MeV are equal.  We
assume a flat crosss-section in the range $30~{\rm MeV} <
\varepsilon' \lesssim 150$~MeV of $6.6\times 10^{-27}$~cm$^2$ and
$1.7\times 10^{-27}$~cm$^2$ respectively for iron and oxygen
nuclei, satisfying above condition.  This more acccurate
cross-section affects the photodisintegration rate in case the
photon spectrum is very hard as we will see later.  For soft
photon spectrum equation (2) is adequate and often a delta
function approximation $\sigma_A (\varepsilon') \sim \sigma_{0,A}
\Delta_{GDR} \delta (\varepsilon' - \varepsilon_0)$ may provide
order of magnitude estimate. }

The photopion cross-section formula for Delta resonance production is
well known and is parametrized by M\"ucke et al. (2000) as
\begin{equation}
\sigma_{\Delta} (\varepsilon') =
\frac{\sigma_{0,\Delta} s^2\Delta_\pi^2} {\varepsilon^{'2} \left[
\left(m_\Delta^2 c^4 -s\right)^2 + s\Delta_\pi^2 \right]},
\end{equation}
where $s=m_p^2c^4+2 \varepsilon' m_pc^2$ is the center-of-mass
energy, $\sigma_{0,\pi} = 3.11\times 10^{-29}$~cm$^2$ and the peak
cross-section is $\sim 4.12\times 10^{-28}$~cm$^2$ at
$\varepsilon' \approx 0.3$~GeV.  The width of the resonance is
$\Delta_\pi = 0.11$~GeV.  The photonuclear cross section is
$\sigma_{\pi}\simeq A^{2/3}\sigma_\Delta$.  Comparing this cross
section to that of photo-disintegration for an iron nucleus with
$A=56$ or Oxygen nucleus with $A=16$, we can see that the
photopion cross section is smaller than the photodisintegration
cross section.  As a result, the photodisintegration rate is
typically higher than the photopion rate.

The rate for spallation process due to collision with other nucleons
in the sources may be estimated as
\begin{equation}
t_{sp}^{-1}=
\sigma_{sp} n_N \beta c,
\end{equation}
where $\sigma_{sp}=5\times10^{-26}A^{2/3}~ {\rm cm^{-2}}$ is the
cross section for spallation of a nucleus with atomic number $A$,
$n_N$ is the number density of nucleus in the sources and $\beta$
is the velocity of the UHE nucleons.  As we show below, even for
internal shocks which have the highest nucleon density, this
spallation effect is less important than photo-disintegration.

Below we will discuss the photodisintegration process in all
three possible UHE nuclei sources, i.e. GRB internal shocks, GRB
external shocks and hypernova remnant blast waves, respectively.

\section {GRB internal shocks}
\label{sec:int}

Due to the variable injection at the base, internal collisions
occurr within the unsteady plasma jet which develop into internal
shocks. The prompt gamma-ray burst emission is supposed to result
from the non-thermal emission of electrons accelerated in these
shocks. The two main processes through which the heavy nuclei
could be disintegrated in internal shocks are photo-disintegration
or photopion interaction by the X-ray photons in prompt emission,
and spallation due to collisions with other nucleons in the
relativistic jet.

\subsection{Nucleus photo-disintegration due to prompt  X-rays }

The GRB prompt photon spectrum is well fitted in the BATSE range
(10 KeV - 3 MeV) by a combination of two power-laws,
$n(\varepsilon)\propto \varepsilon^{-\beta}$, with different
values of $\beta$ at low and high energy (Band et al. 1993).  The
break energy $\varepsilon_{b,obs}$ in the observer frame is
typically $\varepsilon_{b,obs}=\Gamma \varepsilon_b\sim 1 {\rm
MeV}$, with $\beta\simeq 1$ at energies below the break and
$\beta\simeq 2$ above the break, where $\Gamma$ is the bulk
Lorentz factor of the relativistic flow that develops internal
shocks and $\varepsilon_b$ is the break photon energy in the flow
rest-frame. At  low energy, in the framework of the internal shock
model, there could be another spectral break, i.e., the synchrotron
self-absorption break  $\varepsilon_{ssa}$.

Thus the photon spectrum, in the comoving frame, is
\begin{equation}
n(\varepsilon)= \left\{
\begin{array}{l}
n_b(\varepsilon/\varepsilon_b)^{-2} \,\,\,\,\,\,\,\, \varepsilon>\varepsilon_b\\
n_b(\varepsilon/\varepsilon_b)^{-1} \,\,\,\,\,\,\,\, \varepsilon_{ssa}<\varepsilon<\varepsilon_b\\
n_b(\varepsilon_{ssa}/\varepsilon_b)^{-1}(\varepsilon/\varepsilon_{ssa})
\,\,\,\, \varepsilon<\varepsilon_{ssa}
\end{array} \right.  \;.
\end{equation}
where $n_b\equiv n(\varepsilon_b)$ is the photon number density at
the break energy in the comoving frame of the wind.

Using this photon spectrum, the inner integral in Eq.(1) gives
$\int_{\varepsilon/2\gamma}^{\infty}n(\varepsilon)\varepsilon^{-2}d\varepsilon=
\frac{n_b}{(1+\beta)\varepsilon_b}(\frac{\varepsilon}{2\gamma\varepsilon_b})^{-(1+\beta)}
$. The energy density of photons in the wind rest-frame in the
energy range of BATSE is $U_\gamma=n_b\varepsilon_b^2(1/2+{\rm
ln}(\varepsilon_M/\varepsilon_b))\simeq  2n_b\varepsilon_b^2$,
where $\varepsilon_M=3 {\rm MeV}$ is the upper energy range of
BATSE. This energy density is related to the observed photon
luminosity of GRBs, whose typical value is $L_\gamma
\sim10^{51}{\rm erg s^{-1}}$, by $L_\gamma=4\pi R_{in}^2 \Gamma^2
c U_\gamma$, where $R_{in}$ is the radius of the emitting region
of these photons.

The first integral of Eq.(1) can be done numerically for the exact
 cross sections in Eq.(2) nd (3). For the sake of an analytic treatment,
we approximate the cross section as being mainly contributed by
the resonance peak and find that
\begin{equation}
t_{dis}^{-1}= \left\{
\begin{array}{l}
(\frac{U_\gamma}{2\varepsilon_b})c\sigma_0(\frac{{\Delta_{\rm
GDR}}}{\varepsilon_0})
(\frac{2\gamma\varepsilon_b}{\varepsilon_0}),\,\,\,\,\,\,
2\gamma\varepsilon_b\le\varepsilon_0 \\
(\frac{U_\gamma}{2\varepsilon_b})c\sigma_0(\frac{{\Delta_{\rm
GDR}}}{\varepsilon_0}),\,\,\,\,\,\,\,\,\,\,
2\gamma\varepsilon_{ssa}\le\varepsilon_0\le2\gamma\varepsilon_b\\
(\frac{U_\gamma}{2\varepsilon_b})c\sigma_0(\frac{{\Delta_{\rm
GDR}}}{\varepsilon_0}) \kappa^{-2} \times(2{\rm ln}\kappa+1), \,
2\gamma\varepsilon_{ssa}\ge\varepsilon_0
\end{array} \right.  \;.
\end{equation}
where $\kappa\equiv2\gamma\varepsilon_{ssa}/\varepsilon_0$. The
observer-frame energy of the nucleus that interacts preferentially
with photons with energy $\varepsilon_b$ and $\varepsilon_{ssa}$
are, respectively
\begin{equation}
\left\{
\begin{array}{l}
E_b=10^{17}(A/56)\Gamma_{2.5}^2(\varepsilon_{b,obs}/1{\rm
MeV})^{-1} {\rm eV} \\
E_{ssa}=10^{20}(A/56)\Gamma_{2.5}^2(\varepsilon_{ssa,obs}/1{\rm
KeV})^{-1} {\rm eV}
\end{array} \right.
\end{equation}
where $\Gamma \sim 300 \Gamma_{2.5}$ is a typical GRB bulk Lorentz factor.

Now we can obtain the  time $t_{dis}$ of an nucleus of energy $E$
in the observer frame: for $E_b<E<E_{ssa}$
\begin{equation}
t_{\rm dis}=0.035 (\frac{A}{56})^{-1.21}L_{\gamma,51}^{-1} R_{in,
13}^2 \Gamma_{2.5}(\frac{\varepsilon_{b,obs}}{1 {\rm MeV}}) {\rm
s},
\end{equation}
and for $E>E_{ssa}$,
\begin{equation}
t_{\rm dis}\simeq0.035 (\frac{A}{56})^{-1.21}L_{\gamma,51}^{-1}
R_{in, 13}^2 \Gamma_{2.5}(\frac{\varepsilon_{b,obs}}{1 {\rm MeV}})
(\frac{E}{E_{ssa}})^2{\rm s}.
\end{equation}
The dynamic time of internal shock is
\begin{equation}
t_{\rm dyn} =R_{in}/c\Gamma=1 R_{in,13}\Gamma^{-1}_{2.5} {\rm s}.
\end{equation}
By comparing Eq.(8) or Eq.(9) with Eq.(10), we can see that the
heavy nuclei can survive when either
\begin{equation}
R_{in,13} \Gamma_{2.5}^2\ga 30
L_{\gamma,51}(\frac{\varepsilon_{b,obs}}{1 {\rm
MeV}})^{-1}(\frac{A}{56})^{1.21},
\end{equation}
corresponding to $t_{\rm dis}\ga t_{\rm dyn}$, or
\begin{equation}
\varepsilon_{ssa,obs}\ga5
R_{13}^{-1/2}\Gamma_{2.5}E_{20}^{-1}(\frac{A}{56})^{1.6}L_{\gamma,51}^{1/2}(\frac{\varepsilon_{b,obs}}{1
{\rm MeV}})^{-1/2} {\rm KeV},
\end{equation}
in the case of $t_{\rm dis}\la t_{\rm dyn}$.

For comparison, the acceleration time of UHE nuclei with nuclear
charge $Z$ in the internal shock is
\begin{equation}
t_{\rm acc}=\frac{\alpha E}{Z\Gamma e B c}=1.5\times10^{-2} \alpha
E_{20}\epsilon_{B,-1}^{-1/2}
R_{in,13}L_{k,52}^{-1/2}(\frac{Z}{26})^{-1} {\rm s},
\end{equation}
where $\alpha\sim$ few, describing the ratio between the
acceleration time and Larmor time, $B=(\frac{8\pi \epsilon_B
L_k}{4\pi R^2 \Gamma^2 c})^{1/2}=8\times10^4
\epsilon_{B,-1}^{1/2}L_{k,52}^{1/2}R_{13}^{-1}\Gamma_{2.5}^{-1}
{\rm G}$ is the comoving frame magnetic field in internal shocks.
The synchrotron loss time for UHE nuclei is
\begin{equation}
t_{\rm syn}=\frac{6\pi m_p^4 c^3\Gamma}{\sigma_{\rm T} m_e^2 E
B^2}(\frac{A}{Z})^4\simeq
30\epsilon_{B,-1}^{-1}L_{k,52}^{-1}R_{in,13}^2\Gamma_{2.5}^3
E_{20}^{-1}{\rm s},
\end{equation}
which is a factor of $(A/Z)^4\simeq16$ longer than that of
protons.

By equating $t_{\rm acc}$ with $t_{\rm dyn}$ or $t_{\rm syn}$, we
get the maximum energies of accelerated nuclei for these two cases
respectively,
\begin{equation}
\left\{
\begin{array}{l}
\varepsilon_{max}=6\times10^{21} \alpha^{-1} \epsilon_{B,-1}^{1/2}L_{k,52}^{1/2}\Gamma_{2.5}^{-1} \left(\frac{Z}{26}\right) {\rm eV}\\
\varepsilon_{max}=4\times10^{21}
\epsilon_{B,-1}^{-1/4}L_{k,52}^{-1/4}R_{in, 13}^{1/2}
\Gamma_{2.5}^{-3/2}\left(\frac{Z}{26}\right)^{1/2} {\rm eV}
\end{array} \right.
\end{equation}

{ We compare these timescales in Fig.1 for two different sets of
parameters, i.e. $R=10^{13}{\rm cm}$, $\Gamma=10^{2.5}$ (top
panel) and $R=10^{14}{\rm cm}$, $\Gamma=10^{3}$ (bottom panel).}
We can see that: 1) the iron nucleus can be accelerated to
energies $\ga10^{20}{\rm eV}$, since in both cases, $t_{\rm
acc}<t_{\rm dyn}<t_{\rm syn}$ for $E=10^{20}{\rm eV}$; 2)when the
internal shock radius and/or the bulk Lorentz factor are
relatively large, the photo-disintegration time of iron nuclei is
longer than the dynamic time, as seen in the bottom panel of
Fig.1. Note that the photo-disintegration time and the photopion
interaction times in these figures are calculated numerically
according to Eq.(1) with exact cross sections.

For smaller internal shock radii (Fig.1, top panel), the
photo-disintegration effect becomes important if the photon
spectrum $n(\varepsilon)\sim\varepsilon^{-1}$  extends to $\la {1
\rm KeV}$. { This is consistent with the result in Anchordoqui et
al. (2007b).} However, if there is a self-absorption break at
several KeV in the photon spectrum that leads to a drop in the
number density of the photons  with which the UHE nucleus mainly
interacts, the photo-disintegration effect will be suppressed, as
can be seen from the blue dashed lines, where a synchrotron
self-absorption break at 5 KeV in the photon spectrum has been
assumed. This is possible for a small internal shock radius, since
a smaller internal shock radius might lead to higher synchrotron
self-absorption break $\varepsilon_{ssa}$. The value of
$\varepsilon_{ssa}$ also depends on the dissipation model as well
as the radiation mechanism for prompt emission (e.g. Panaitescu \&
\Meszaros 2000; Rees \& \Meszaros 2005; Pe'er et al. 2005; Pe'er
\& Zhang 2006), which is, however, largely uncertain at present.
For the simple optically thin internal shock model,
$\varepsilon_{ssa}$ is estimated to be
\begin{equation}
\varepsilon_{ssa,
obs}\sim1L_{k,52}\epsilon_{e,-1}^{-1/3}R_{13}^{-2/3}\Gamma_{2.5}^{1/3}
{\rm KeV}
\end{equation}

For UHE intermediate-mass nuclei, such as O nuclei, the
constraints for survival, Eqs.(11) and (12), are much looser.
Thus, for small internal dissipation or shock radii,
intermediate-mass nuclei are much more likely to survive
photo-disintegration. This can be also seen in Fig. 2, where all
the timescales are calculated for O nuclei with $Z=8$ and $A=16$.

\subsection{Nuclear spallation due to collisions}
\label{sec:spallation}

An UHE nucleus can also spallate due to collision with other
nucleons in internal shocks. The nucleon density of a relativistic
flow with kinetic energy luminosity $L_k=10^{52}{\rm erg s^{-1}}$
at internal shock radius $R_{in}$ is $n_N=L_k/(4\pi  R_{in}^2
\Gamma^2 m_N c^3)=1.8\times10^{12} L_{k,52}
R_{in,13}^{-2}\Gamma_{2.5}^{-2} {\rm cm^{-3}}$. So the spallation
time  is
\begin{equation}
t_{sp}=\frac{1}{n_N\sigma_{sp}c}=25 L_{k,52}^{-1}
R_{in,13}^2\Gamma_{2.5}^2 (\frac{A}{56})^{-2/3} {\rm s}.
\end{equation}
Comparing this with the dynamic time, we find that the heavy
nucleus can survive if
\begin{equation}
R_{in,13} \Gamma_{2.5}^3\ga 0.04L_{k,52}(\frac{A}{56})^{2/3}.
\end{equation}
This condition is much easier to satisfy, compared to the more
stringent photo-disintegration constraint of Eq.(11).

\section{GRB external shocks}
\label{sec:ext}

As the relativistic GRB jets expand and sweep up the surrounding
material,  external shocks develop. The accelerated electrons in
the external shock produce the observed afterglow emission. The
external shock has also been suggested to  be able to account for
UHECRs. The early deceleration phase, when the jet converts about
half of its energy to the swept-up material, is the most effective
phase for the external shock to accelerate particles, since the
shocked material has the largest  Lorentz factor at that time. The
early external shock produces the early x-ray afterglow, whose
average luminosity observed by {\em Swift}  is about
$L_x\simeq10^{48}{\rm erg s^{-1}}$ during the first  $\sim10- 100
\,{\rm s}$ (Nousek et al. 2006). In this external shock scenario,
the accelerated UHE nuclei with energy $E\ga10^{19}$ eV
preferentially interact with these x-ray photons, if the Lorentz
factor is typically $10^2\la\Gamma\la 10^{3}$.

Taking the early X-ray afterglow spectrum in the fast-cooling
regime (e.g. \Meszaros 2006) with $F_\nu\sim \nu^{-1}$,
corresponding to photon spectral index $\beta=2$, we can obtain
the photo-disintegration rate of a nucleus moving with Lorentz
factor $\gamma_A$, i.e.
\begin{equation}
t_{dis}^{-1}=\frac{4}{3}c \sigma_0 (\frac{\Delta_{\rm
GDR}}{\varepsilon'_0})(\frac{\gamma_A U_x}{\kappa
\varepsilon'_0}),
\end{equation}
where $U_x$ is the comoving-frame energy density of X-ray
afterglow photons and $\kappa={\rm
ln(\epsilon_{X,M}/\epsilon_{X,m})}\simeq 3$, $\epsilon_{X,M}$ and
$\epsilon_{X,m}$ being the upper and lower end of {\em Swift} XRT
energy threshold. The energy density is related to the luminosity
by $L_x=4\pi R_{ex}^2\Gamma^2 c U_x$, where $R_{ex}$ is the radius
of the external shock at the end of the free expansion phase of
the ejecta. Now we can obtain the photo-disintegration energy loss
time of a cosmic ray nucleus of energy $E$ propagating in the
early afterglow photons
\begin{equation}
t_{\rm dis}=3\times10^{6} R_{ex,17}^2 \Gamma_{2.5}^3 L_{x,48}^{-1}
E_{20}^{-1}(\frac{A}{56})^{-0.21} {\rm s}.
\end{equation}
Comparing this with the dynamic time in the comoving frame of external
shocks  $t_{\rm dyn}=R_{ex}/(2c\Gamma)=10^{4}
R_{ex,17}\Gamma_{2.5}^{-1} {\rm s}$, we can see that the heavy
nucleus can survive when
\begin{equation}
R_{ex,17}\Gamma_{2.5}^4\ga 3\times10^{-3} L_{x,48} E_{20}
(\frac{A}{56})^{0.21}.
\end{equation}
For a constant density medium environment , this gives an upper
limit on the number density of the medium
\begin{equation}
n\la2.5\times10^6E_{k,53}\Gamma_{2.5}^{10}L_{x,48}^{-3}E_{20}^{-3}(A/56)^{-0.63}
{\rm cm^{-3}},
\end{equation}
while for a stellar wind environment with a density profile
$\rho=\left(\frac{\dot{M}}{4\pi v_w}\right)r^{-2}=5\times10^{11}
A_{\star}r^{-2} {\rm g cm^{-1}}$, this gives constrains on the
mass-loss rate parameter
\begin{equation}
A_{\star}\la1.3E_{k,53}\Gamma_{2.5}^{-2}L_{x,48}^{-1}E_{20}^{-1}(A/56)^{-0.21},
\end{equation}
where $A_{\star}=1$ corresponds to $\dot{M}=10^{-5} M_\odot {\rm
yr^{-1}}$ and $v_w=1000 {\rm km s^{-1}}$.

After the initial free expansion phase, the external shock starts
to decelerate as more and more material is swept-up.  The shock
radius evolves with time as $R_{ex}(t)=4\times10^{17}
E_{k,53}^{1/4}n_0^{-1/4}t_d^{1/4} {\rm cm}$ for constant density
medium and
$R_{ex}(t)=2\times10^{18}E_{k,53}^{1/2}A_{\star,-1}^{-1/2}t_{\rm
d}^{-1/2} {\rm cm}$ for wind medium, where $t_d$ is time in units
of days. Assuming that the magnetic field energy density acquires
a fraction $\epsilon_B=0.1$ of the internal energy, the magnetic
field evolves as
$B=1\epsilon_{B,-1}^{1/2}E_{k,53}^{1/8}n_0^{3/8}t_d^{-3/8}{\rm G}$
for a constant density medium and $B=0.07
\epsilon_{B,-1}^{1/2}E_{k,53}^{-1/4}A_{\star,-1}^{3/4}t_{\rm
d}^{-3/4}{\rm G}$ for wind medium. Since the synchrotron loss is
typically unimportant for UHE nuclei in external shocks,  the
maximum energy of particles accelerated in external shocks is
obtained by equating $t_{\rm acc}$ with $t_{\rm dyn}$, which gives
\begin{equation}
\begin{array}{l}
\varepsilon_{max}=ZeBR_{ex}(t) \\
=\left\{ \begin{array}{l}
3\times10^{21}(\frac{Z}{26})\epsilon_{B,-1}^{1/2}E_{k,53}^{3/8}n_0^{1/8}t_{d}^{-1/8} {\rm eV}   \,\,\,(\rm ISM)\\
10^{21}(\frac{Z}{26})\epsilon_{B,-1}^{1/2}E_{k,53}^{1/4}A_{\star,-1}^{1/4}t_{d}^{-1/4}
{\rm eV} \,\,\,(\rm wind)\\  \end{array} \right .
\end{array}
\end{equation}
This shows that the maximum energy of accelerated particles by
external shock decreases rather slowly with time, so that even
weeks to months after the burst, heavy nuclei can be still
accelerated to UHE energies by afterglow shocks, { even if the
flux may decrease with time.}

The optical depth $\tau$ for photo-disintegration of these nuclei
evolves with time as
\begin{equation}
\tau=\frac{ t_{\rm dyn}}{t_{\rm dis}}\sim L_{x}\Gamma^{-4}
R_{ex,t}^{-1} \sim \left\{
\begin{array} {l}
t^{1/4} \,\,\,\, (\rm ISM) \\
t^{-1/2} \,\, (\rm wind)
\end{array} \right.
\end{equation}
where an X-ray afterglow luminosity $L_{x}\sim t^{-1}$ has been
assumed. The slowly increasing optical depth or  decreasing
optical depth indicates that the accelerated heavy nuclei by GRB
external shocks can survive  for a relatively long time. We plot
in Fig. 3 the evolution of photo-disintegration optical depth and
the maximum energy of accelerated particles  for iron nucleus (top
 panel) and Oxygen nucleus (bottom panel) respectively.

\section{Hypernova remnant blast wave}
\label{sec:hn}

The observations of the radio afterglow of the hypernova SN1998bw
showed that about $10^{50}~{\rm erg}$ of kinetic energy were
released in the form of a mildly relativistic ejecta (Kulkarni, et
al. 1998; Chevalier \& Li 1999). The interpretation of the X-ray
afterglow also favors a mildly-relativistic ejecta component
(Waxman 2004b). A recently detected strong thermal X-ray emission
component in another sub-energetic burst (GRB060218), associated
with SN2006aj, may also be associated with a mildly-relativistic
supernova shock breakout, in which the mildly relativistic
supernova ejecta has an energy $\gtrsim10^{49}~{\rm erg}$ (Campana
et al. 2006). Due to the large supernova explosion energy and the
much lower than typical GRB energy, attempts have been made to
ascribe the prompt gamma-ray emission to the shock from the mildly
relativistic ejecta as it breaks out through the hypernova
progenitor's outer envelope (Woosley et al. 1999; Tan et al.
2001), although  a generally accepted conclusion has not yet been
reached. We have used the term semi-relativistic hypernovae to
denote such supernovae exhibiting a mildly relativistic ejecta
component, seen in association with GRBs. Such high-velocity
ejecta are a key ingredient of the hypernova model for UHECRs,
since the low-velocity bulk ejecta is not able to accelerate
particles to such high energies. Based on the shock breakout
scenario, we have suggested  that there might be a continuous
distribution of ejecta energy in velocity, i.e.
$E_k(\Gamma\beta)\simeq 3\times10^{52}
(\Gamma\beta/0.1)^{-\alpha}$ with $\alpha\simeq 2$, where $\beta$
is the ejecta velocity in units of speed of light (Wang et al
2007).

As the hypernova ejecta expand, they transfer their energy to the
swept-up stellar wind, and external shocks develop, which
accelerate wind particles to ultra-high energies. Since the
ejecta has a velocity distribution profile, the leading edge
higher-velocity ejecta decelerate the earliest  and then the
lower-velocity ejecta decelerate progressively. The maximum energy
of accelerated particles is related to  the ejecta velocity
$\Gamma\beta$ by
\begin{equation}
\begin{array}{ll}
\varepsilon_{\rm max}\simeq Z e BR\beta \\
= 1.3\times10^{20} (\frac{Z}{26})
\epsilon_{B,-1}^{1/2}\left(\frac{\Gamma\beta}{0.5}\right)^{2}A_{\star}^{1/2}
{\rm eV}.
\end{array}
\end{equation}
Although higher velocity ejecta can accelerate particles to higher
energies, the kinetic energy in ejecta of $\Gamma\beta> 0.5$ are
too low to account for the UHECR flux (Wang et al. 2007), due to
the steep distribution of energy $E_k\propto (\Gamma\beta)^{-2}$.
For ejecta with a velocity $\Gamma\beta$, the free expansion
phase before deceleration sets in lasts for a time
\begin{equation}
t_f=1300 (\Gamma\beta/0.5)^{-5} A_{\star}^{-1} {\rm
days}
\end{equation}
and the radius of the ejecta at this time is
\begin{equation}
r_f=1.7\times10^{18}(\Gamma\beta/0.5)^{-4} A_{\star}^{-1} {\rm
cm}.
\end{equation}
There are two photon sources which could cause
photo-disintegration of heavy nuclei: one is provided by hypernova
thermal photons from radioactive elements of the hypernova ejecta,
and another is the synchrotron photons from the hypernova remnant
blast wave. For the first mechanism, we  use the luminosity  of
SN1998bw as a representative. At time $t\simeq1000$ days after the
burst, the optical luminosity of SN1998bw drops to the level of
about $L_{\rm HE}\sim10^{39}{\rm erg s^{-1}}$ (Sollerman et al.
2002). A nucleus of energy $E=10^{20}$ eV interacts with target
photons with energy $\varepsilon_t\ga0.01 ({A}/{56}) E_{20}^{-1}
{\rm eV}$. A rough estimate of the optical depth of
photo-disintegration of heavy nuclei due to hypernova thermal
photons is
\begin{equation}
\begin{array}{ll}
\tau\la \sigma_0(\frac{L_{\rm HE}}{4\pi r_f^2 c \varepsilon_{\rm
HE}})(\frac{r_f}{\eta}) \\
=3\times10^{-5} L_{\rm HE,39}
r_{f,18}^{-1} (\frac{\varepsilon_{\rm HE}}{ 1{\rm eV}})^{-1},
\end{array}
\end{equation}
where $\eta\simeq 4$ is the compression ratio of the hypernova
external shock and $\varepsilon_{\rm HE}\simeq 1 {\rm eV}$ is the
characteristic energy of hypernova thermal photons.  We can see
that thermal photons are so sparse that they have a negligible
effect on the photo-disintegration of UHE nuclei.

The synchrotron emission from the stellar wind shocked by the
hypernova remnant could be brighter than the hypernova emission
itself at the late stages of the hypernova, e.g. $t_f\sim 1300{\rm
days}$ after the burst in our case, since the hypernova luminosity
drops rather quickly after the peak. We estimate the luminosity
from the shocked wind at the time when the $\Gamma\beta=0.5$
ejecta begins to decelerate. The total number of the swept-up wind
particles is $N_{e}= \left(\frac{\dot{M}}{ v_w}\right)
r_f=6.6\times10^{54} (t_f/1300 \,{\rm d})^{-4/5}A_{\star}^{-4/5}$,
where ${\rm d}$ denotes time in days. Assuming that the magnetic
field energy density is amplified to a fraction $\epsilon_B$ of
the shock internal energy, the magnetic field is
$B=0.015(\Gamma\beta/0.5)^{5}A_{\star}^{3/2}{\rm G}=
0.015(t_f/1300 \,{\rm d})^{-1}A_{\star}^{1/2}{\rm G}$. Assuming a
power-law energy distribution for accelerated electrons
$dn_e/d\gamma_e\propto \gamma_e^{-p}$ with $p\simeq 2$, one
obtains the characteristic frequency $\nu_m=7\times10^5
\epsilon_{e,-1}^2 (t_f/1300{\rm d})^{-9/5}A_{\star}^{-3/10}{\rm
Hz}$, where $\epsilon_e$ is the fraction of the shock energy that
goes into electrons. High energy electrons will cool in the
magnetic field and cause a break in the electron energy
distribution. The characteristic synchrotron frequency
corresponding to this break is $\nu_c=4\times10^{13}(t_f/1300{\rm
d})A_{\star}^{-3/2}{\rm Hz}$. The luminosity at the frequency
$\nu_m$ is $L_{\nu_m}=N_{e,m} P(\gamma_{e,m})\simeq 4\times
10^{36} \epsilon_{e,-1}^2\epsilon_{B,-1}(t_f/1300{\rm
d})^{-2}A_{\star} \, {\rm erg s^{-1}}$. For $p\simeq 2$, the
luminosity at frequency $\nu$ is
$L_{\nu}=L_{\nu_m}(\nu/\nu_m)^{1/2}$ for $\nu_m<\nu<\nu_c$, while
for $\nu>\nu_c$, $L_{\nu}\simeq
L_{\nu_c}=L_{\nu_m}(\nu_c/\nu_m)^{1/2}$. The luminosity at energy
$\varepsilon_t=0.01 ({A}/{56}) E_{20}^{-1} {\rm eV}$,
corresponding to the energy of the photons with which the UHE
nuclei of energy $E$ interact at the resonance peak,  is $L_{\rm
syn}=10^{40}\epsilon_{e,-1}\epsilon_{B,-1}^{3/4}(t_f/1300{\rm
d})^{-1.1}A_{\star}^{1.15}{\rm erg s^{-1}}$. So we see that for
wind parameters $A_{\star}=1$, the synchrotron luminosity does
exceed the hypernova luminosity at the time $\sim10^3$
days{\footnote{Unlike in SN2006aj, the inferred stellar wind for
SN1998bw is much weaker, with $A_{\star}=0.04-0.1$, which may
explain why the optical emission of SN1998bw, detected at
$\sim1000$days after the burst, is still contributed by the
hypernova emission }}. Once we know the synchrotron luminosity of
a hypernova remnant and the photon spectral index ($\beta=3/2$),
we can,  using Eq.(1), get the optical depth of
photo-disintegration of heavy nuclei due to such synchrotron
photons, i.e.
\begin{equation}
\tau\simeq 8\times10^{-3}
\epsilon_{e,-1}\epsilon_{B,-1}^{3/4}A_{\star}^{1.35}E_{20}^{1/2}({\frac{A}{56}})^{0.71}(\frac{t_f}{1300{\rm
d}})^{-1.9} .
\end{equation}
Using $\tau\la1$, we find that, after a time
\begin{equation}
t_f\ga 100
\epsilon_{e,-1}^{0.53}\epsilon_{B,-1}^{0.4}A_{\star}^{0.7}E_{20}^{0.26}(\frac{A}{56})^{0.37}
{\rm d},
\end{equation}
the nucleus can survive in the hypernova synchrotron photon
environment. Note that the free expansion time for ejecta with
$\Gamma\beta\la0.5$ (Eq.27) is longer than this time. The maximum
energy of accelerated particles, Eq.(26), depends on the time
$t_f$ as
\begin{equation}
\varepsilon_{\rm max}\simeq 1.3\times10^{20}
\left(\frac{Z}{26}\right) \epsilon_{B,-1}^{1/2}
\left(\frac{t_f}{1300{\rm d}}\right)^{-2/5} A_{\star}^{1/10} {\rm
eV}.
\end{equation}
The evolution of the optical depth for photo-disintegration and
the maximum energy of accelerated  nuclei are shown in Fig.4.

\section{Discussion and Conclusions}
\label{sec:disc}

The possible presence of heavy nuclei in UHECRs raises interesting
questions: what is the origin of these nuclei? can these nuclei
survive in the sources where they get accelerated? In this paper,
we have endeavored to address these questions for two proposed
UHECR sources discussed in the literature, namely GRBs and
hypernovae. For GRBs, both internal shocks and external shocks
have been suggested to be able to accelerate particles to UHE
energies, and in this paper we have considered the role of both of
these in the context of UHECR heavy nuclei. We have sketched out
some possible { mechanisms for the presence of} heavy nuclei in
these scenarios. For GRB internal shocks, we suggest that the
nuclei are entrained from the progenitor stellar core, e.g. the Fe
core, the O core, etc., during the stage when the jets arising
from accretion in the innermost collapsing core, are making their
way out through the star.  {Since instabilities favoring
entrainment are predominant mainly during the initial stellar
crossing phase of the jet, a jet composition with a substantial
fraction of heavy nuclei may only be expected in long bursts with
observed $\gamma$-ray durations $t_\gamma\lesssim 10-15(1+z)$ s,
while longer burst jets would be expected to consist mainly of
protons. }  For the GRB external shock and hypernova source
scenarios, the nuclei may be the heavy elements present in the
stellar wind of Wolf-Rayet stars, which are thought to be the
progenitors of GRBs and hypernovae. The stellar wind of Wolf-Rayet
stars, especially the WO, WC sub-types stars, is heavily enriched
with intermediate mass nuclei, such as O, C, etc.

{After they escape from their sources (see e. g. Dermer 2007b),
UHECR nuclei can be subject to photo-disintegration also in
intergalactic space, before arriving at Earth.  An UHE heavy
nucleus with an energy $\ga10^{19}{\rm eV}$ will mainly collide
with cosmic infrared background (CIB) photons. Using new
constraints on CIB data, it has been found that UHE iron nuclei
with energy $\la10^{20}{\rm eV}$ have a mean free path of $\ga
300{\rm Mpc}$ and that this mean free path increases  as the
energy of the nucleus decreases (Hooper et al. 2007; Stecker \&
Salamon 1999). Lighter nuclei have relatively shorter mean free
path (Hooper et al. 2007). Thus, UHE nuclei can in principle
originate from sources at cosmological distance, such as GRBs and
hypernovae. The main question appears to be whether they survive
the environment of their original sources. For this reason, most
of the present paper has been devoted to a quantitative  study of
the survival of UHE nuclei in the above mentioned sources.  We
find that: }

i) In GRB internal shocks, heavy nuclei can survive in the sources
if the internal shock radius and/or the Lorentz factor of the
relativistic jets are relatively large, as given by Eq.(11).  Thus
one might expect acceleration of Fe nuclei to be more favored in
bursts with smoother, longer variability timescale light-curves.
For a smaller internal shock radius, the photodisintegration
process due to prompt X-ray photons may become optically thick.
However, if a synchrotron self-absorption break is present above
several keV in the photon spectrum, the reduced number density of
X-ray photons will lower the photodisintegration optical depth
accordingly. In general, compared to heavy nuclei, the UHE
intermediate-mass nuclei find it easier to survive
photo-disintegration in internal (and external) shocks.

ii) In GRB external shocks, due to  the much larger dissipation
radii compared to internal shocks, UHE nuclei can easily survive
in the sources. We have also calculated the evolution of the
photo-disintegration optical depth during the afterglow
phase for both an ISM external medium and a stellar wind medium.
The results show that UHE nuclei can survive in the afterglow
shock for a relatively long time  in both cases.

iii) In the hypernova remnant acceleration scenario, UHE nuclei
can survive in the sources, except in the early short period of
time ($\la 100 {\rm days}$ for typical parameters, see Eq.(31)).
In this early short period, however, only a very small amount of
ejecta energy has been converted to UHECRs, so their contribution
to the UHECR flux is negligible. Most of the UHECRs originate from
the blast wave ejecta with $\Gamma\beta\la0.5$, which  decelerate
after $\sim 1000 {\rm days}$, typically. UHE nuclei accelerated
during this time are safe from photo-disintegration.

{ Earlier  calculations on the photo-disintegration of UHE nuclei
in GRB internal shocks by Anchordoqui et al. (2007b) consider only
one specific set of typical parameters for the internal shock
radius and the relativistic Lorentz factor. We have improved this
by considering the whole parameter space and working out the
constraints  on the physical conditions under which UHE heavy
nuclei  can survive in internal shocks. We have also considered
the effect of a self-absorption break in the prompt emission on
the photo-disintegration problem. Besides internal shocks, we have
explored the photo-disintegration problem of UHE nuclei in GRB
external shocks and hypernova remnants. }

In summary, we have suggested possible scenarios for the injection
of heavy nuclei into the acceleration zones of extragalactic
sources such as GRB and hypernovae. We found that, if  heavy
nuclei are accelerated in these sources, they will survive the
threat of photo-disintegration under fairly general conditions for
the case of GRB external shocks, and for hypernovae. They could
survive also in GRB internal shocks, if the latter occur at
relatively large radii and/or the bulk Lorentz factors are large.
On the other hand, for small shock radii and/or smaller bulk
Lorentz factors, a pure proton UHECR composition would be favored
if the self-absorption break in the photon spectrum is not high.
Since the instability-induced entrainment process of heavy nuclei
in internal shocks is currently not well-known, we stress that the
expected fraction of heavy nuclei injected into the internal shock
acceleration process is uncertain. A significant fraction is
plausible, but if the entrainment is inefficient, we would in any
case expect a proton-dominated composition in GRB internal shock
models. In GRB external shocks and hypernova shocks, on the other
hand, the abundance of heavy nuclei is dependent on the external
medium or the stellar outer envelope and wind enrichment fraction.

{ As we were completing our work, the Pierre Auger Collaboration
(2007) reported a plausible correlation between UHECR at energies
above $6\times10^{19}$ eV, assuming that they are protons, and
AGNs within 75 Mpc selected from a particular catalog. Although
this result is statistically significant at the 2.8$\sigma$ level,
as the authors themselves stress, it does not rule out a possible
origin in sources, e.g. galaxies, which are distributed in a
similar manner as these AGNs.  There is an unresolved tension
between the above Auger spatial correlation analysis suggesting
protons, and previous Auger results on maximum shower elongations
$X_{max}$ suggesting a significant heavy element component for
UHECR in the same energy range. { The AGN hypothesis for the
UHECRs has been questioned in the analysis of Gorbunov et al.
(2007).} In our present paper, we have investigated the conditions
under which UHECR originating from GRB or hypernovae could contain
a significant fraction of heavy nuclei, as well as the conditions
under which these UHECR would be expected to be mainly protons. If
any of the observed UHECR were heavy nuclei, these could reach the
Earth from distances beyond 100 Mpc, and the number of galaxies
which can host a GRB or hypernova increases significantly with
distance. Within the proton-inspired 3.2 degree circle containing
an AGN in the Auger correlation analysis, there would be many more
galaxies which could host a GRB or a hypernova. In the case of
heavy nuclei, the deflection angles are larger (e.g.  for Oxygen
with energy $\ga 6\times10^{19}$ eV it is $\la 16$ degree, Sommers
2007), which would contain even more galaxies. While a study of
the angular correlations is beyond the scope of this paper, our
results are compatible with and relevant for both the Auger
elongation studies and spatial correlation studies, providing
constraints which can be used in future analyses.}

{\acknowledgments We would like to thank Charles Dermer for
helpful comments and Zhuo Li for pointing out that the
self-absorption break may have an effect on the
photo-disintegration.  This work is supported in part by
NSF AST 0307376, NASA NAG5-13286, and  the National Natural Science
Foundation of China under grants 10403002 and 10221001, and the
Foundation for the Authors of National Excellent Doctoral
Dissertations of China (for X.Y.W.).}


\newpage
\begin{figure*}
\centering 
\includegraphics[width=12cm]{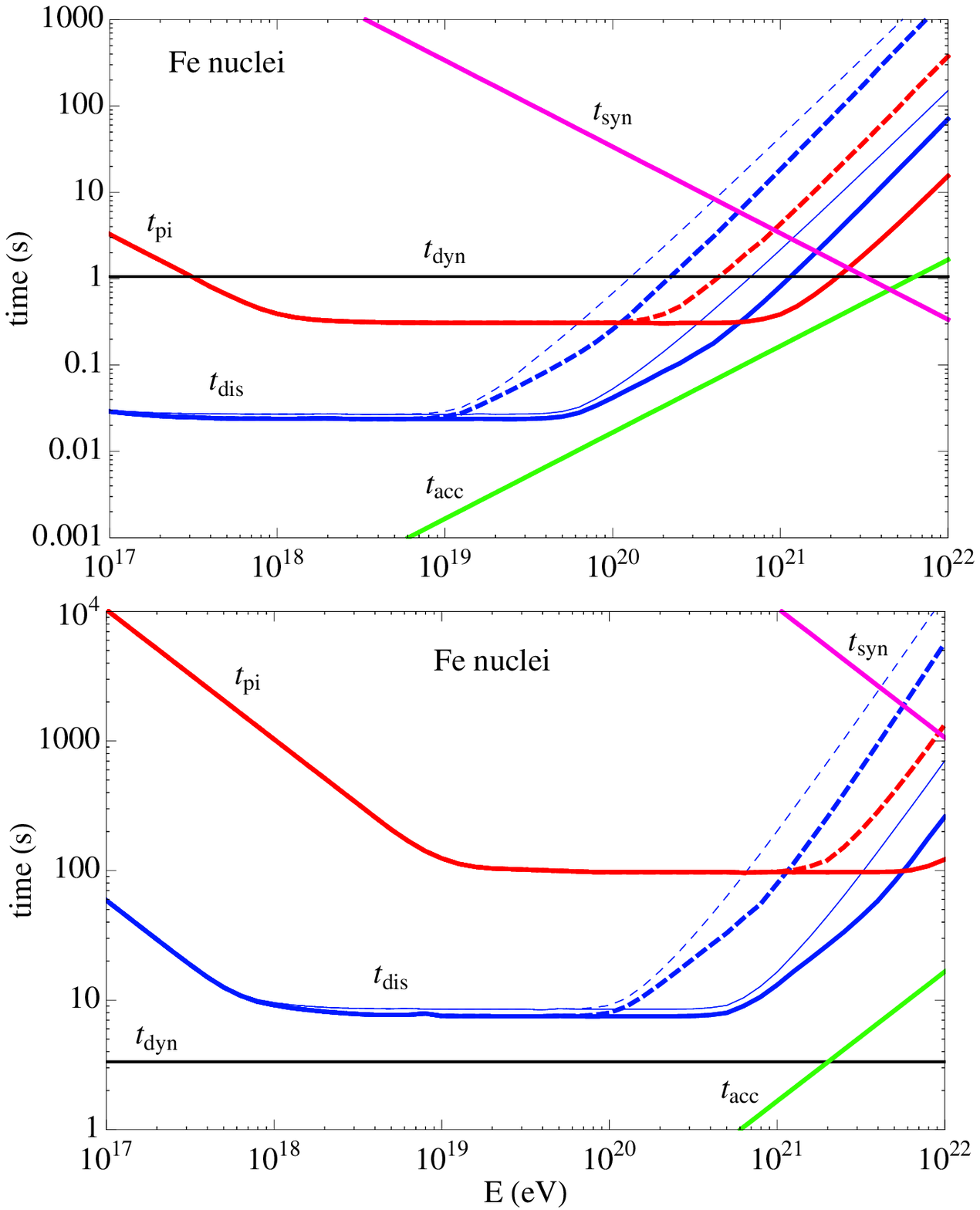}
\caption{\small Top
panel: Timescales in the internal shock (or in general, small
dissipation radius) scenario, comparing the disintegration time
$t_{\rm dis}$, the photopion interaction time $t_\pi$, the dynamic
time $t_{\rm dyn}$, the particle acceleration time $t_{\rm acc}$
and the synchrotron loss time $t_{\rm syn}$ for iron nuclei at
different energy $E$. The disintegration time (thick blue lines)
and photopion interaction (thick red lines) times are calculated
numerically by using exact cross-sections, i.e., equation (2) and
multi-nucleon emission cross-section for photodisintegration and
equation (3) for photopion.  The thin blue lines corresponds to
photodisintegration time using equation (2) only, i.e., ignoring
the multi-nucleon emission cross-section.  For the
photodisintegration and photopion times, the solid line and dashed
lines are for a synchrotron self-absorption break
$\varepsilon_{ssa,obs}$ at 1 KeV and 5 KeV respectively. The
dissipation radius and jet Lorentz factor used in this plot are
$R_{\rm in}=10^{13}~{\rm cm}$ and $\Gamma=10^{2.5}$. Note that the
maximum energy of an accelerated iron nucleus is
$\simeq3\times10^{21}~{\rm eV}$ from $t_{\rm acc}=t_{\rm syn}$ at
this radius.  Bottom panel: The same as above but for a larger
dissipation radius, $R_{\rm in}=10^{14}~{\rm cm}$ and a larger jet
Lorentz factor $\Gamma=10^3$.  The maximum energy of an
accelerated iron nucleus is $\simeq2\times10^{21}~{\rm eV}$ from
$t_{\rm acc}=t_{\rm dyn}$.  The acceleration time has been roughly
approximated as the Larmor time, i.e. $\alpha=1$ in both plots.}
\end{figure*}

\begin{figure*}
\centering 
\includegraphics[width=12cm]{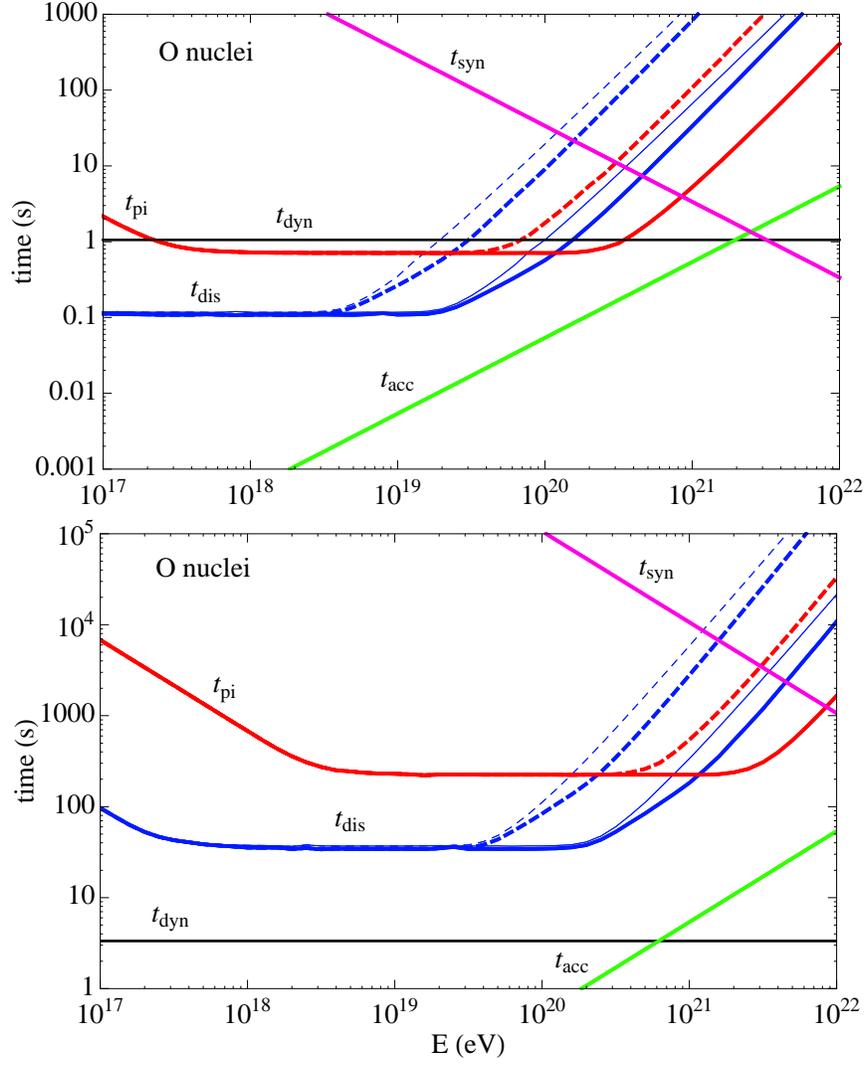}
\caption{The
same as Fig.1, but for UHE Oxygen nuclei. The maximum energy of an
accelerated O nucleus is $\simeq 10^{21}~{\rm eV}$ (top panel,
$R_{\rm in}=10^{13}~{\rm cm}$) and $\simeq 6\times10^{20}~{\rm
eV}$ (bottom panel, $R_{\rm in}=10^{14}~{\rm cm}$), both from the
condition $t_{\rm acc}=t_{\rm dyn}$.  Note that the O nuclei above
$few\times 10^{19}$~eV may survive in the GRB internal shocks
taking place even at smaller radii.}
\end{figure*}

\begin{figure*}
\centering 
\includegraphics[width=12cm]{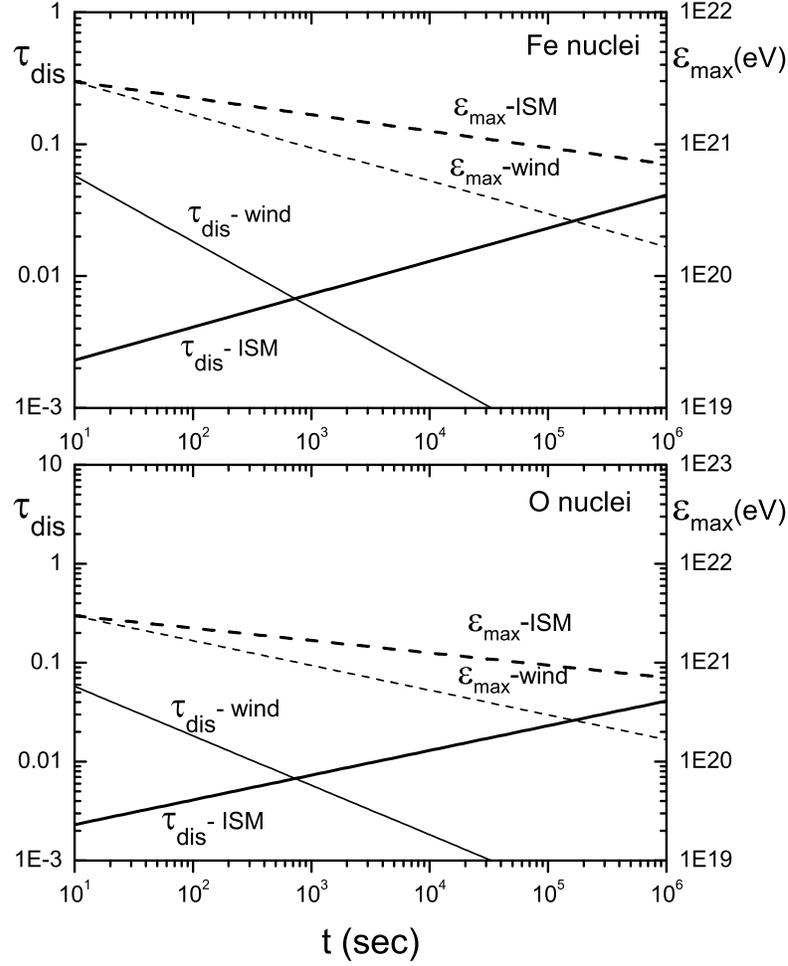}
\caption{External shocks scenario. In the top panel, dashed lines show the
evolution of the maximum energy of an accelerated iron nucleus for
an external shock moving into a constant density medium (thick
dashed line) and into a wind medium (thin dashed line), while the
solid lines show the evolution of the photo-disintegration optical
depth of an iron nucleus of energy $E=10^{20}{\rm eV}$ in external
shocks moving into the constant density medium (thick solid line)
and the wind medium (thin solid line). The  parameters used for
the constant density medium case and the wind medium case are,
respectively, \{$n=1{\rm cm^{-3}}$, $E=10^{53}{\rm erg}$,
$\Gamma=10^{2.5}$ , $\epsilon_B=0.1$ \} and \{$A_{\star}=0.1$,
$E=10^{53}{\rm erg}$, $\Gamma=10^{2.5}$, $\epsilon_B=0.1$ \}.
Bottom panel: the same as the top panel, but for UHE Oxygen
nuclei.}
\end{figure*}

\begin{figure*}
\centering
\includegraphics[width=12cm]{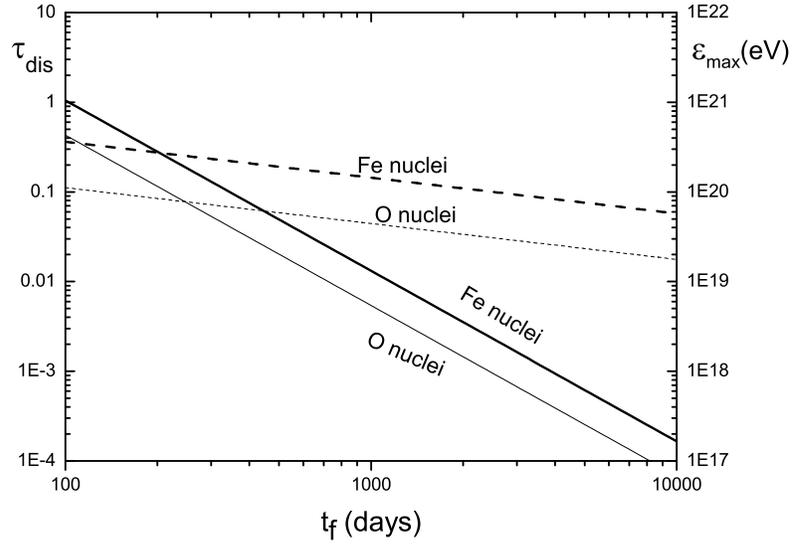}
\caption{Hypernova acceleration scenario. Dashed lines: evolution of the
maximum energy of  an accelerated iron nucleus (thick dashed line)
and an Oxygen nucleus (thin dashed line) in the hypernova remnant;
Solid line: evolution of the photo-disintegration optical depth
of an iron nucleus (thick solid line) and an Oxygen nucleus (thin
solid line) of energy $E=10^{20}{\rm eV}$. The parameters used are
$A_{\star}=1$, $\epsilon_e=0.1$ and $\epsilon_B=0.1$.}
\end{figure*}

\end{document}